\documentclass[sigconf]{acmart}
\AtBeginDocument{%
  }

\copyrightyear{2026}
\acmYear{2026}
\setcopyright{cc}
\setcctype{by}
\acmConference[IUI '26]{31st International Conference on Intelligent User Interfaces}{March 23--26, 2026}{Paphos, Cyprus}
\acmBooktitle{31st International Conference on Intelligent User Interfaces (IUI '26), March 23--26, 2026, Paphos, Cyprus}
\acmPrice{}
\acmDOI{10.1145/3742413.3789134}
\acmISBN{979-8-4007-1984-4/2026/03}
\usepackage{xcolor}

\newcommand{\xhdr}[1]{\vspace{0em}\noindent{{\bf #1.}}}

\newcommand{\hide}[1]{}

\usepackage{float}
\usepackage{hyperref}
\usepackage{cleveref}   % for cref

\begin{document}

\title{Improving Human Verification of LLM Reasoning through Interactive Explanation Interfaces}
% Towards Interactive Alternatives to Traditional Chain-of-Thought Reasoning in Large Language Models}


\author{Runtao Zhou}
\affiliation{%
  \institution{University of Virginia}
  \city{Charlottesville}
  \state{Virginia}
  \country{USA}}
\email{uar6nw@virginia.edu}

\author{Giang Nguyen}
\affiliation{%
  \institution{Auburn University}
  \state{Alabama}
  \city{Auburn}
  \country{USA}
}
\email{nguyengiangbkhn@gmail.com}

\author{Nikita Kharya}
\affiliation{%
 \institution{Independent Researcher}
 \city{Charlottesville}
 \state{Virginia}
 \country{USA}
 }
\email{nikitakharya09@gmail.com}

\author{Anh Totti Nguyen}
\affiliation{%
  \institution{Auburn University}
  \city{Auburn}
  \state{Alabama}
  \country{USA}}
\email{anh.ng8@gmail.com}

\author{Chirag Agarwal}
\affiliation{%
  \institution{University of Virginia}
  \city{Charlottesville}
  \state{Virginia}
  \country{USA}}
\email{qze3wn@virginia.edu}

\renewcommand{\shortauthors}{Trovato et al.}

\begin{abstract}

\looseness=-1 The reasoning capabilities of Large Language Models (LLMs) have led to their increasing employment in several critical applications, particularly education, where they support problem-solving, tutoring, and personalized study. 
While there are a plethora of works showing the effectiveness of LLMs in generating step-by-step solutions through chain-of-thought (CoT) reasoning on reasoning benchmarks, little is understood about whether the generated CoT is helpful for end-users in improving their ability to comprehend mathematical reasoning problems and detect errors/hallucinations in LLM-generated solutions.
To address this gap and contribute to understanding how reasoning can improve human-AI interaction, we present three new interactive reasoning interfaces: interactive CoT (iCoT), interactive Program-of-Thought (iPoT), and interactive Graph (iGraph), and a novel framework that generates the LLM's reasoning from traditional CoT to alternative, interactive formats.
Across 125 participants, we found that interactive interfaces significantly improved performance. Specifically, the iGraph interface yielded the highest clarity and error detection rate ($85.6\%$), followed by iPoT ($82.5\%$), iCoT ($80.6\%$), all outperforming standard CoT ($73.5\%$).
Interactive interfaces also led to faster response times, where participants using iGraph were fastest ($57.9$ secs), compared to iCoT and iPoT ($60$ secs), and the standard CoT baseline ($64.7$ secs). 
Furthermore, participants preferred the iGraph reasoning interface, citing its superior ability to enable users to follow the LLM's reasoning process. 
We discuss the implications of these results and provide recommendations for the future design of reasoning models. 
The code and interfaces for this project can be found here: \href{https://github.com/Runtaozhou/Interactive-CoT}{https://github.com/Runtaozhou/Interactive-CoT}
\end{abstract}





\maketitle

\section{Introduction}
\label{sec:intro}
% \chirag{THIS NEEDS A COMPLETE NEW PASS}
% \chirag{Write the high-level story:
% \begin{enumerate}
%     \item What is the context -- what are the benefits of CoT -- why are we targeting CoT -- what are the RQs that we are hypothesizing in this work -- contributions
%     \item Points to consider: However, arguably, these are being developed without embracing the full spectrum of end-user needs.
% \end{enumerate}
% }
Large language models (LLMs) are now woven into high-stakes, real-world workflows, including education, where they assist with problem solving, tutoring, and personalized study~\cite{kojima2022large, wang2024large,forootani2025survey}. A key driver of their recent success on complex tasks is chain-of-thought (CoT) prompting~\cite{wang-etal-2023-towards}: eliciting step-by-step intermediate reasoning often yields dramatic accuracy gains on arithmetic, commonsense, and symbolic reasoning benchmarks, with state-of-the-art performance on GSM8K and related math datasets. However, despite these advancements, the current format for presenting LLM explanations is far from optimal~\cite{agarwal2024faithfulness}.
% Large Language Models (LLMs) have emerged as powerful tools for solving complex reasoning tasks\cite{kojima2022large, wang2024large}. It demonstrates remarkable performance across domains such as mathematics, programming, and logical inference\cite{forootani2025survey}. Models such as GPT, Claude, and PaLM have shown the ability to produce step-by-step reasoning through techniques like chain-of-thought (CoT) prompting by leveraging vast amounts of training data and advanced architectures\cite{yu2023towards, kandpal2025position}. These detailed explanations provide insights into model reasoning and help users understand intermediate steps leading to a final answer\cite{wang-etal-2023-towards}. However, despite these advancements, the current format for presenting LLM explanations is far from optimal\cite{agarwal2024faithfulness}.

\begin{figure*}
    \centering
    \includegraphics[width=\textwidth]{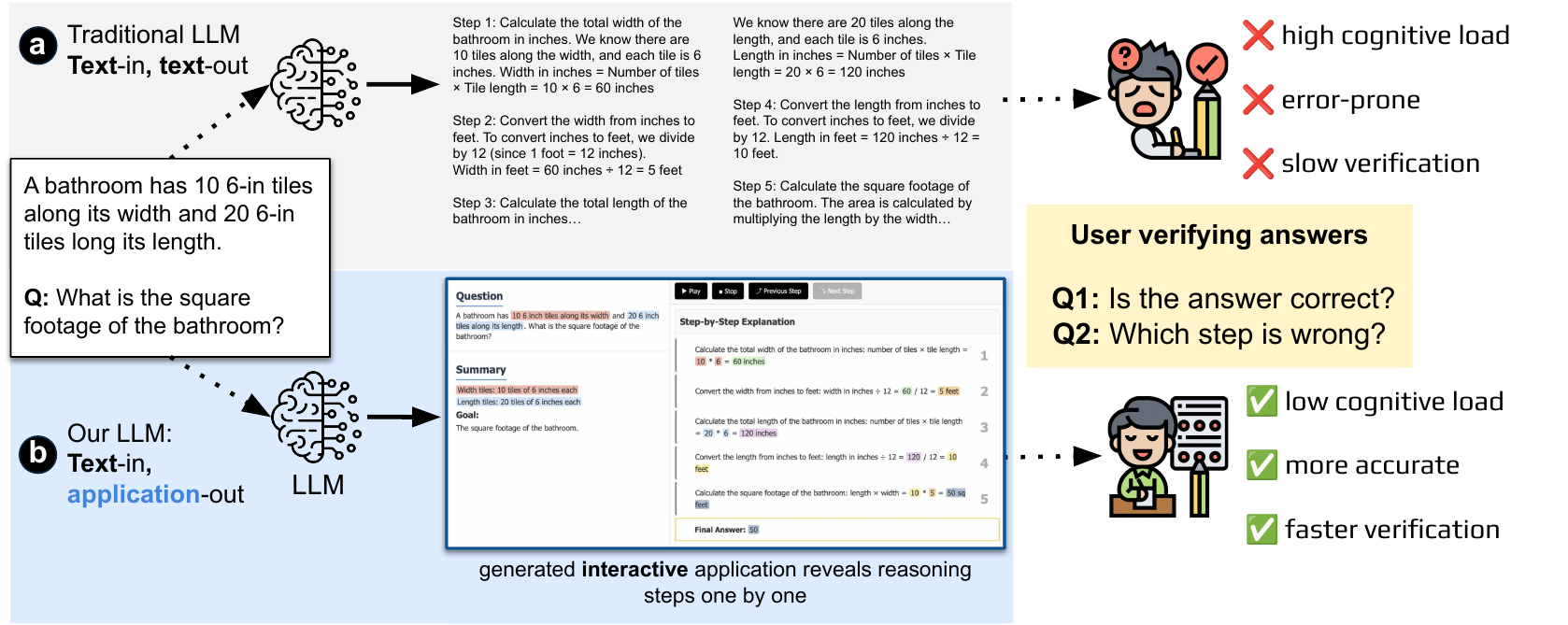}
    \caption{
    Given a GSM8K question, LLMs typically provide step-by-step reasoning followed by the final answer.
    However, such output presentation is often static and long, posing a higher cognitive load to users and leading to slower and more erroneous answer verification. 
    In contrast, we prompt LLMs to \emph{generate an interactive HTML/JavaScript application} wrapped around the reasoning.
    This interface enables users to verify the reasoning more efficiently via tools of (a) navigation buttons (inspired by those in common IDEs) and (b) colored highlights \cite{nguyen2025hot}.
    }
    \label{fig:teaser}
\end{figure*}

Despite these advances, it remains unclear whether text-only CoT explanations actually help end users—students, teachers, or non-expert verifiers—understand and check mathematical reasoning. While these outputs contain rich information, their format introduces significant challenges for human interpretation~\cite{kunz2024properties, hedlin2025got, turpin2023language}. In practice, CoT often appears as a long ``\textit{wall of text}'' that can be time-consuming, obscure structure, increase cognitive load, and make error detection harder precisely when it matters. For example, when an explanation contains a subtle mistake in one intermediate step, detecting that error often requires carefully parsing multiple sentences and mentally reconstructing the reasoning flow~\cite{mishra2025investigating}. More broadly, the explainable AI (XAI) literature shows that explanation techniques are frequently designed around what researchers can produce rather than what end users need; users instead report wanting practically useful information that improves collaboration, trust calibration, and feedback loops with AI systems\cite{miller2019explanation, abdul2018trends, kaur2020interpreting}. This gap motivates rethinking how reasoning is surfaced to end-users in contexts like education, where clarity, verification, and engagement are essential.

% Most LLM-generated explanations are presented as long, unstructured blocks of text. While these outputs contain rich information, their format introduces significant challenges for human interpretation\cite{kunz2024properties, hedlin2025got, turpin2023language}. Users must read through a dense “wall of text” to identify critical reasoning steps, verify correctness, and understand logical dependencies\cite{si2023large}. This process can be time-consuming, cognitively demanding, and prone to error. For example, when an explanation contains a subtle mistake in one intermediate step, detecting that error often requires carefully parsing multiple sentences and mentally reconstructing the reasoning flow \cite{mishra2025investigating}. As a result, users may overlook incorrect steps or incorrectly assume that an explanation is valid simply because it appears thorough and coherent\cite{kamoi2024evaluating}. These challenges are particularly concerning in high-stakes settings—such as education, finance, law, and healthcare—where incorrect reasoning can lead to serious consequences~\cite{tavasoli2025responsible,wiggins2022opportunities}.

% Therefore, improving the interpretability and usability of LLM explanations is an important open problem. 
Prior work has demonstrated that the design of LLM explanations, which includes factors such as visual structure, interactivity, and information hierarchy, can significantly influence users' trust and understanding~\cite{yang2020visual, alicioglu2022survey, jiang2023graphologue,nguyen2024interpretable}. However, most existing research on CoT focuses on improving the reasoning process within the model rather than optimizing the presentation of the reasoning~\cite{wang2024chain,wang2022towards}. As a result, LLM explanations remain largely static, text-heavy, and difficult to navigate~\cite{zhao2024explainability}. This disconnect raises critical questions: \textit{How can explanation formats be redesigned to reduce cognitive load and enable users to effectively detect errors and understand LLM-generated reasoning steps?}
% Can alternative presentation styles enable users to more effectively detect errors and understand reasoning steps?} and support accurate verification?}

\xhdr{Present work} To address these questions, we propose an interactive explanation interface, an approach that explores interactive and structured alternatives to traditional text-based reasoning outputs as shown in Fig.\ref{fig:teaser}. Rather than presenting explanations as static and dense paragraphs, we introduce three novel formats designed to enhance clarity and interpretability: i) interactive CoT (iCoT), a format that preserves the text-based explanation but introduces features such as color-coded highlights for key steps, a clear step-by-step breakdown, and a concise summary for quick reference, ii) interactive graph (iGraph), a format where reasoning steps are represented as nodes connected by edges in a left-to-right flow diagram, allowing users to easily identify the logical progression of steps and visually detect errors or inconsistencies and iii) interactive Program-of-Thought (iPoT), a code-based format that presents reasoning as a sequence of pseudo-code lines. Each line represents a logical operation or variable update, showing how intermediate values lead to the final solution.
% \begin{enumerate}
%     \item \textbf{ Enhanced Natural Language Format:} This format preserves the text-based explanation but introduces features such as color-coded highlights for key steps, a clear step-by-step breakdown, and a concise summary for quick reference.
%     \item \textbf{Code-Based Format:} This format is inspired by structured programming syntax and organizes reasoning steps like a code snippet. By leveraging computational thinking patterns, it makes the reasoning process clearer and easier to follow.
%     \item \textbf{Graphical Interface:} In this format, reasoning steps are represented as nodes connected by edges in a left-to-right flow diagram. This visual representation allows users to easily identify the logical progression of steps and visually detect errors or inconsistencies.
% \end{enumerate}

To evaluate the effectiveness of these formats, we conducted a controlled user study using problems from GSM8K, a multi-step mathematical reasoning benchmark. 
% Explanations were generated using Claude 3.7 Sonnet. This will ensure consistency and quality across conditions. 
Each participant was randomly assigned to one of the four explanation formats, including a traditional text baseline, and asked to review ten LLM-generated (using Claude 3.7 Sonnet) explanations.
% Of these, one explanation was fully correct, while nine contained errors deliberately introduced in specific steps. 
Participants were tasked with determining whether each explanation was correct or incorrect. For incorrect explanations, the user has to identify the exact step where the error occurred. This experimental design allows us to measure two key performance metrics: verification accuracy, which reflects participants' ability to correctly identify errors and locate faulty steps, and time per question, which indicates the efficiency of error detection. 
% Together, these measures provide insights into how different presentation formats affect user performance. 
Finally, we performed a post-experiment study, where we took users' subjective feedback and measured the user engagement during the experiment. Our study aims to answer three central research questions: 

\textbf{RQ1)} Do interactive explanation formats improve verification accuracy compared to traditional text outputs?

\textbf{RQ2)} How do these formats influence the time and effort required to verify reasoning?

\textbf{RQ3)} Do users prefer interactive explanation formats over traditional text-based outputs? 

\looseness=-1 By systematically comparing these formats, we aim to uncover design principles that enhance the interpretability of LLM-generated explanations. Our work contributes not only a practical evaluation of interactive explanation designs but also broader implications for building trustworthy systems. When users can efficiently and accurately verify reasoning steps, they are better equipped to detect model errors, avoid over-reliance on incorrect outputs, and make informed decisions. These findings have potential applications in educational tools, auditing systems, and human-AI collaboration platforms, where understanding and validating reasoning is essential. In summary, this paper makes three primary contributions: i) we introduce interactive explanation frameworks, an approach that explores three alternative explanation formats (iCoT, iPoT, and iGraph) to improve the interpretability and usability of LLM reasoning; ii) our findings, from a controlled user study, show the effectiveness of iCoT, iPoT, and iGraph over traditional chain-of-thought on verification accuracy and time efficiency; and iii) we discuss design implications for developing interactive and interpretable AI systems that support user trust, error detection, and effective human-AI collaboration.

% \begin{enumerate}
%     \item We introduce Interactive CoT, an approach that explores three alternative explanation formats—enhanced natural language, code-like representation, and graphical visualization—to improve interpretability and usability of LLM reasoning.
%     \item We present findings from a controlled user study comparing these formats to a traditional text baseline on verification accuracy and time efficiency.
%     \item We discuss design implications for developing interactive and interpretable AI systems that support user trust, error detection, and effective human-AI collaboration.
% \end{enumerate}

\section {Related Work}
Here, we outline the purpose of CoT reasoning in LLMs through the lens of improving user understanding and the role of interactive presentation in supporting comprehension and engagement within educational contexts.

\looseness=-1\xhdr{LLMs in Education with Chain-of-Thought (CoT) Prompting} A key prompting technique for LLMs is CoT prompting, which elicits the LLM's reasoning steps and improves its performance on multi-step tasks~\cite{zhang2022automatic}. A recent study shows that few-shot CoT dramatically boosts arithmetic, commonsense, and symbolic reasoning accuracy~\cite{wei2022chain,kojima2022large}. In educational settings, however, how explanations are presented matters. Despite clear opportunities for tutoring and feedback,~\citet{khosravi2022explainable} highlights persistent concerns about transparency, interpretability, and over-reliance in LLMs. Beyond transparency, CoT reasoning tends to be quite lengthy, which can increase cognitive load and slow down the user's comprehension~\cite{nayab2024concise}. Given the concerns about the interpretability and cognitive demands of LLM-generated reasoning, researchers have begun to explore how reasoning quality and user understanding can be evaluated more systematically. A growing body of work has adopted verification accuracy, which indicates the users' ability to correctly judge whether a model's reasoning is valid, as a key indicator of comprehension in LLM explanations~\cite{narayanan2018humans, linder2021level, nourani2020don, lage2019human,nguyen2024interpretable}. Further, a recent study argues that explanations are only useful when they enable users to verify predictions step-by-step, rather than simply accepting plausible reasoning at face value~\cite{fok2024search}.In our study, we aim to address this question by measuring how effectively people can detect reasoning errors across different explanation formats, allowing us to assess not only the quality of the reasoning itself but also how various interface designs influence users' ability to verify and understand it.

\looseness=-1\xhdr{Interactivity in LLM Explanations} Recent studies have emphasized the importance of interactivity in AI explanations to improve deeper user engagement and understanding\cite{kim2020improved,guesmi2024interactive,nguyen2024allowing}.~\citet{teso2023leveraging} provides a comprehensive overview of interactive machine learning, which shows that explanations are most effective when users can query, edit, or refine them rather than passively receive them. They argue that interaction transforms explanations from static outputs into dynamic learning tools that encourage users' cognitive engagement. Similarly, another work reviews human-centered explainability and finds that interactive explanation interfaces significantly enhance engagement, comprehension, and retention compared to one-way text outputs~\cite{zhang2025human}. Their analysis shows that many existing AI systems remain static and fail to leverage interactive elements that foster user engagement and a deeper understanding of the AI reasoning process.
In educational settings, a study demonstrates that interactive LLM-driven question–answer systems increase learners' focus and cognitive engagement without introducing cognitive load~\cite{ozdel2025examining}. Together, these studies highlight that interactivity in AI explanations not only improves the user's understanding but also keeps the users engaged.

\looseness=-1\xhdr{Transparency and reliance in LLM Explanations} Prior work in human-centered XAI highlights that transparency and reliance are closely related but not equivalent\cite{de2024we}. Recent work claims that transparency emerges from the combined design of models, data, and interfaces, and advocates for simplified and interactive explanations to support user understanding\cite{muralidhar2023elements}. Another work argues that explanations can unintentionally induce over- or under-reliance, and that current evidence on trust calibration remains inconclusive without explicit reliance-aware UI design\cite{de2024we}. Building on top of these insights, a recent work on selective transparency via progressive disclosure demonstrates that layered explanations reduce cognitive overload while supporting appropriate reliance in clinical settings\cite{muralidhar2025operationalizing}.These works suggest LLM explanations must be transparent, context-aware, and designed to calibrate reliance rather than maximize information disclosure.

Our work extends these insights by introducing interactive reasoning formats that move beyond passive text outputs. We explore how interactivity can make LLM reasoning more transparent, interpretable, and engaging by allowing users to navigate reasoning steps, inspect structure, and verify logic.

\section{Method}
\label{sec:method}

\subsection{Designing Alternative Response Formats for LLM Reasoning}
The goal of this project is to design and evaluate alternative response formats for LLM-generated reasoning for math problems. While LLMs have demonstrated strong capabilities in producing step-by-step reasoning through CoT prompting, prior research shows that these outputs often are \textbf{lengthy, text-heavy passages} that are difficult for students to parse and verify in practice~\cite{plevris2023chatbots}. Therefore, we investigate whether alternative reasoning formats can enhance comprehension and reduce the cognitive load of users. To address these challenges, we design three alternative interactive reasoning formats that transform the same underlying content into different modes of presentation. We now introduce our three alternative explanation formats alongside the baseline explanation format.
% \noindent\textbf{Interactive coding (iCode)} outputs the solution as Python-style code, where each reasoning step is represented as a line of code accompanied by a descriptive comment. A dynamic ``variable display'' panel updates to show the current values of variables in real time, with consistent color-coding that matches the variables introduced in the problem statement. This design draws inspiration from computational thinking. The coding interface encourages users to track both the logic and the numerical outcomes of each step by explicitly linking code, commentary, and variable state.

\begin{figure*}
    \centering
    \includegraphics[width=\textwidth]{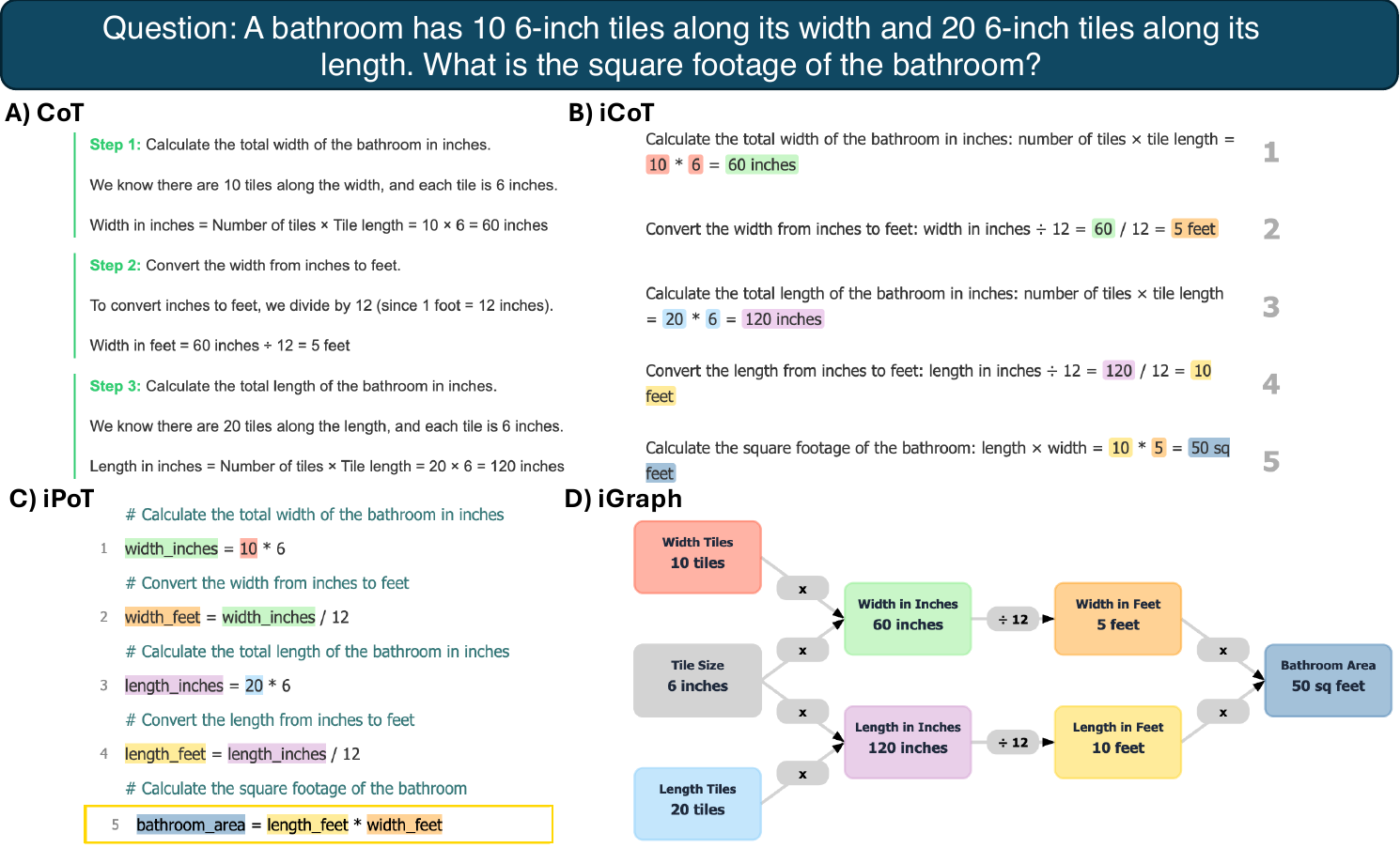}
    \caption{Examples of four explanation formats used in the study: (A) traditional Chain-of-Thought (CoT), (B) interactive Chain-of-Thought (iCoT), (C) interactive Program-of-Thought (iPoT), and (D) interactive Graph (iGraph). Each format presents the same reasoning steps in a different modality (textual, structured, code-like, or visual). For consistency, all four formats present the same mathematical problem. Complete interfaces for these four example explanation formats are displayed in \cref{appendix:display} (\cref{fig:cot,fig:igraph}).
    }
    \label{fig:interface_display}
\end{figure*}

\looseness=-1\noindent\textbf{Traditional Chain-of-Thought (CoT)} serves as the baseline explanation format. It presents reasoning in a purely text-based form, where the model expresses its thought process step-by-step in natural language. Each step builds upon the previous one and forms a continuous logic flow that leads to the final solution\cite{wei2022chain}. The interface includes three main sections: the problem statement, the model’s step-by-step explanation, and the final answer (see Fig. \ref{fig:interface_display} A). This format represents the standard way large language models naturally generate explanations without additional structural guidance or interactivity.

\noindent\textbf{Interactive Chain-of-Thought (iCoT)} demonstrates a more traditional format but restructures the reasoning into discrete, highlighted segments (see Fig. \ref{fig:interface_display} B). iCoT introduces three key interactive features: (1) a dual-panel design where the left panel summarizes key information and statistics from the question for easy reference; (2) color-coded highlights that emphasize critical information and numbers across both panels, helping users identify and cross-reference key details; and (3) navigation buttons at the top of the interface that display one reasoning step at a time, which allow users to progress through the explanation at their own pace.These features work together with the aim of making it easier for users to detect inconsistencies or mistakes.

\looseness=-1\noindent\textbf{Interactive Program-of-Thought (iPoT)} shows explanations in a structured, code-like format (see Fig. \ref{fig:interface_display} C). Each reasoning step is represented as a line of pseudo-code, where variables are defined, updated, and computed systematically. This format allows users to trace how each intermediate value is derived through reasoning steps. Playback controls enable users to execute the reasoning “step-by-step,” similar to debugging or running a script. This interface design draws inspiration from the “Program of Thoughts Prompting” framework proposed by \citet{chen2022program}, which demonstrated that representing reasoning as executable programs can improve interpretability and consistency in LLMs.

\noindent\textbf{Interactive Graph (iGraph)} renders reasoning as a node-link diagram (see Fig. \ref{fig:interface_display} D). Each node corresponds to either a problem variable, an intermediate value, or a final result. The edges represent operations connecting them. 
% Playback controls buttons on the top of the interface allow users to step forward or backward through the solution, and at each stage, a synchronized textual explanation is presented to reinforce the visual display. 
This approach aims to show the links between steps more clearly and help learners understand the solution through structured visuals.

The three interactive reasoning formats retain the content of a traditional CoT, restructure how it is presented to support verification, and reduce cognitive load. Each interactive explanation format follows a consistent interface template with specific design features to minimize variation in presentation. Next, we outline the design features of our interface to explain their functionality and relevance.

\looseness=-1\xhdr{1) Dual-Panel Layout} All interactive explanations adopt a standardized dual-panel design. The left panel displays the problem statement and summary, while the right panel presents the step-by-step explanation. This separation ensures that users can always reference the original problem context \textbf{without scrolling or losing track} of the reasoning process. 

\xhdr{2) Problem Summarization} A concise problem summary is displayed beneath the problem statement in the left panel. It highlights the key variables along with their numerical values, providing the users a quick reference to check against as they follow each reasoning step.

\xhdr{3) Colored Variables} Variables are consistently color-coded across both panels. Each variable is assigned a unique color that is retained wherever it appears (problem statement, problem summary, or step-by-step explanation). This helps the users trace variables along with their values through the entire reasoning chain.

\xhdr{4) Step-by-Step Display with Controls} The explanation is displayed one step at a time in the right panel(except for the iPoT interface). Users can move forward and backward using playback control buttons. This design choice is intended to reduce cognitive overload and help users focus on the reasoning process incrementally rather than being overwhelmed by the entire solution at once.
% \xhdr{5) \chirag{Should we add about the correct/incorrect feature?}}

\subsection{Dataset Construction}
To evaluate how alternative explanation formats aid users in detecting LLM-generated errors, we construct a dataset derived from GSM8K, a widely used benchmark of grade school math word problems~\cite{cobbe2021training}. GSM8K is chosen because it contains a large and diverse set of multi-step reasoning tasks that resemble problems students might encounter in classroom and tutoring settings. Each GSM8K problem typically involves 3–8 steps of reasoning with natural language explanations, which makes it well-suited for transformation into interactive formats.

\looseness=-1 A central aspect of our design is the controlled injection of errors into the explanations. Rather than relying on naturally occurring model mistakes, following~\citet{li2024evaluating}, we adopt the dataset that they create with the taxonomy of nine error categories: Calculation (CA), Counting (CO), Context Value (CV), Contradictory Step (CS), Missing Step (MS), Hallucination (HA), Unit Conversion (UC), Operator (OP), and Formula Confusion (FC). In this dataset, the correct and erroneous reasoning steps have already been labeled and validated. By systematically using each of these error types, we create a consistent and fine-grained framework that enable us to assess not only whether participants could recognize when an explanation is incorrect, but also whether certain categories of errors are more easily detected across different interface formats \footnote{A detailed analysis of verification accuracy for each individual error category is presented in Appendix. \ref{appendix:error_category}}.
% for evaluating the robustness of explanations. This approach enabled us to assess not only whether participants could recognize when an explanation was incorrect, but also whether certain categories of errors were more easily detected across different interface formats.
For each error type, we randomly sample 50 problems from GSM8K. To ensure balanced difficulty across conditions, we sampled explanations containing relatively similar numbers of reasoning steps (between four to seven steps) within each error category. This sampling strategy ensures that participants encounter comparable task complexity regardless of their assigned interface format. This range is chosen because shorter problems may make error detection trivial, while longer problems risk overburdening participants. Altogether, this process produced 450 erroneous explanations (9 error types × 50 examples each). To balance the dataset and establish a baseline for comparison, we also sample an additional 50 correct explanations from GSM8K. Like the erroneous cases, these problems also contain 4–7 reasoning steps in length and served to evaluate whether participants could correctly recognize valid reasoning when presented in different formats. By combining erroneous and correct cases, we obtain a dataset of 500 explanations in total. 

\subsection{Tagging as an Intermediate Representation}
\looseness=-1 To ensure the interface formats are consistent, we develop an intermediate tagging layer that acts as a structured guideline for problem statements and explanations. Each problem is first encoded into tags for facts, steps, formulas, calculations, and variables. For example, \textcolor{red!75!green}{\textbf{<fact>}} elements capture problem parameters, \textcolor{blue}{\textbf{<step>}} elements describe reasoning in natural language, \textcolor{green!10!orange}{\textbf{<formula>}} captures symbolic relations, \textcolor{violet}{\textbf{<calculation>}} contains explicit numeric operations, and \textcolor{purple}{\textbf{<var>}} defines intermediate or final results. Wrong steps are annotated with \textcolor{red}{\textbf{<wrongstep>}}, which indicates the steps containing an injected error. The \textcolor{green!70!orange}{\textbf{<output>}} tags specify the final solution. This tagging schema allows us to separate independent information. Each tagged problem can be rendered into the four interface formats without altering the underlying logic. Errors are introduced systematically by referencing the \textcolor{red}{\textbf{<wrongstep>}} index, allowing us to maintain strict control over consistency while still enabling interactive elements, like graph animations, code execution, and step playback.

\begin{figure*}
    \centering
    \includegraphics[width=0.72\textwidth]{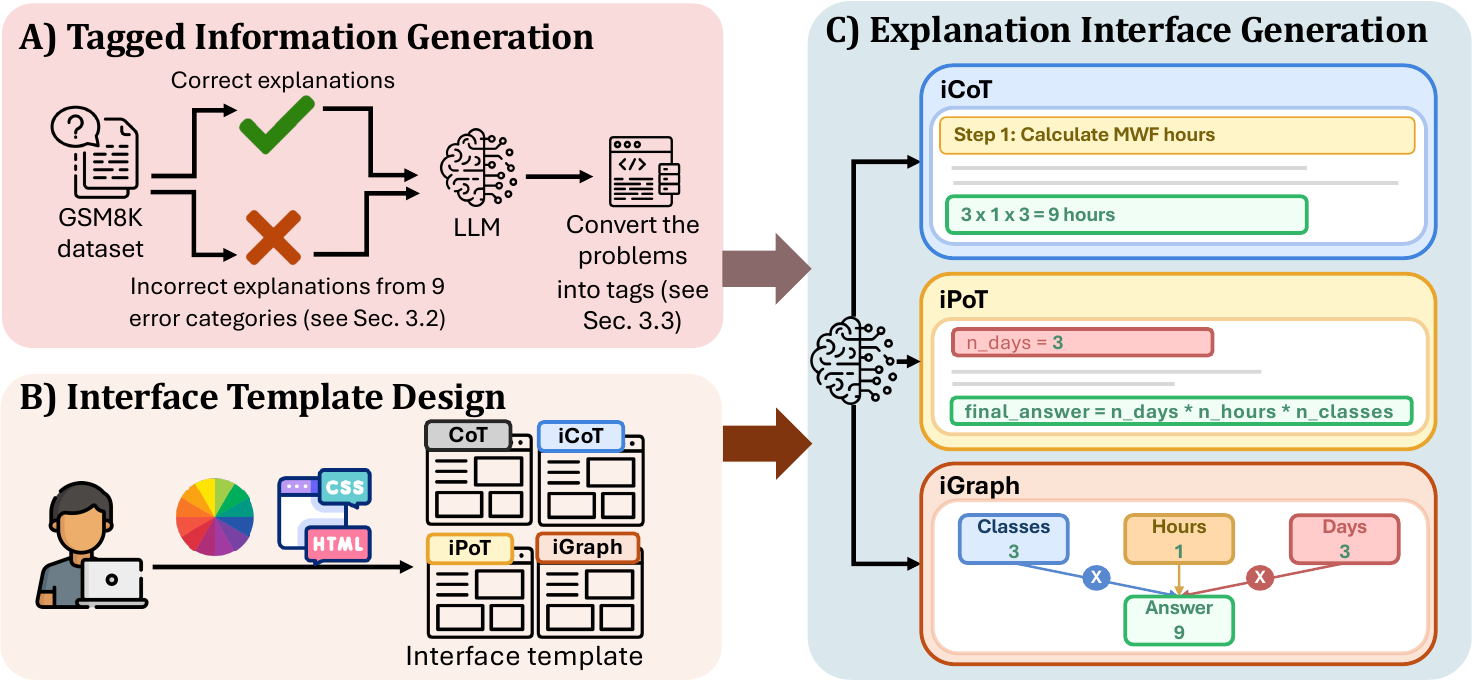}
    \caption{
    % \chirag{Figure one can be the original explanations from each category -- could be a $2 \times 2$ grid; remove claiming iCode as ours and refer to Program of thought; Replace all iCode to PoT; Remove iCode from the figure}
    \looseness=-1\textbf{Our Explanation Generation pipeline} consists of three stages: \textbf{(A)} Tagged Information Generation, where an LLM produces correct and erroneous GSM8K explanations annotated with reasoning tags; \textbf{(B)} Interface Template Design, where standardized HTML/CSS templates ensure consistent structure and interactivity across formats; and \textbf{(C)} Explanation Interface Generation, where tagged data and templates are combined to create interactive explanations in iCoT, iPoT, and iGraph formats for our user study.}
    \vspace{-0.1in}
    \label{fig:overview}
\end{figure*}
\subsection{Interface Templates for Consistency}
While the tagging layer ensures that explanations across formats are generated from the same underlying content, an additional step is required to control for internal variations in raw LLM output. Specifically, we develop interface-specific templates because LLMs can produce responses with differences in phrasing, ordering, or formatting even when prompted identically. These templates define the layout, structure, and presentation rules for each explanation format, ensuring that all outputs within a format adhere to the same design principles. The templates serve two purposes: first, they \textbf{ensure a consistent user experience} within each format by standardizing how reasoning steps, variables, and intermediate values are presented; second, they \textbf{reduce the risk} that uncontrolled LLM variations would bias the evaluation. The templates allow us to isolate the effects of interface design from artifacts of model output by constraining visual and structural differences.

For the traditional CoT interface, the template fixes the structure of step-by-step textual explanations as sequential paragraphs with aligned formulas and calculations. For the iCoT format, the template segments explanations into discrete blocks with color-coded variable highlights and playback controls for navigating step by step. In the iPoT format, the template ensures that each reasoning step appears as a line of pseudocode with comments and that a synchronized variable panel updates consistently across explanations. For the iGraph interface, the template standardizes the node-link arrangement, left-to-right flow, and inclusion of textual support beneath the graph.

\subsection{Data Flow and Interface Generation}
\looseness=-1 Our proposed pipeline from problem input to rendered explanation involves two key stages, as illustrated in Fig.\ref{fig:overview}. First, each math problem and its solution are passed to the LLM (Claude 3.7 Sonnet~\cite{Anthropic} in all our experiments), which produces a tagged intermediate representation. This stage ensures that the reasoning content is captured in a structured format. In the second stage, the tagged representation is combined with the interface-specific templates, where the tags provide the logical content and the templates enforce layout rules and structural consistency. This combined input is then passed to another LLM, which generated the final rendered interface in one of four formats: traditional CoT, iCoT, iPoT, or iGraph.

Our two-stage process allowed us to standardize explanation content through tagging before rendering, thereby eliminating uncontrolled differences that naturally occur from direct LLM outputs. Further, it modularizes the workflow by separating tags from templates, which ensures that the same reasoning can be faithfully expressed across multiple interface modalities. The result is a controlled and reproducible pipeline in which variations in participant performance could be attributed to interface design, rather than inconsistencies in problem or explanation generation.

\section{User Study}
\begin{figure*}
    \centering
    \includegraphics[width=\textwidth]{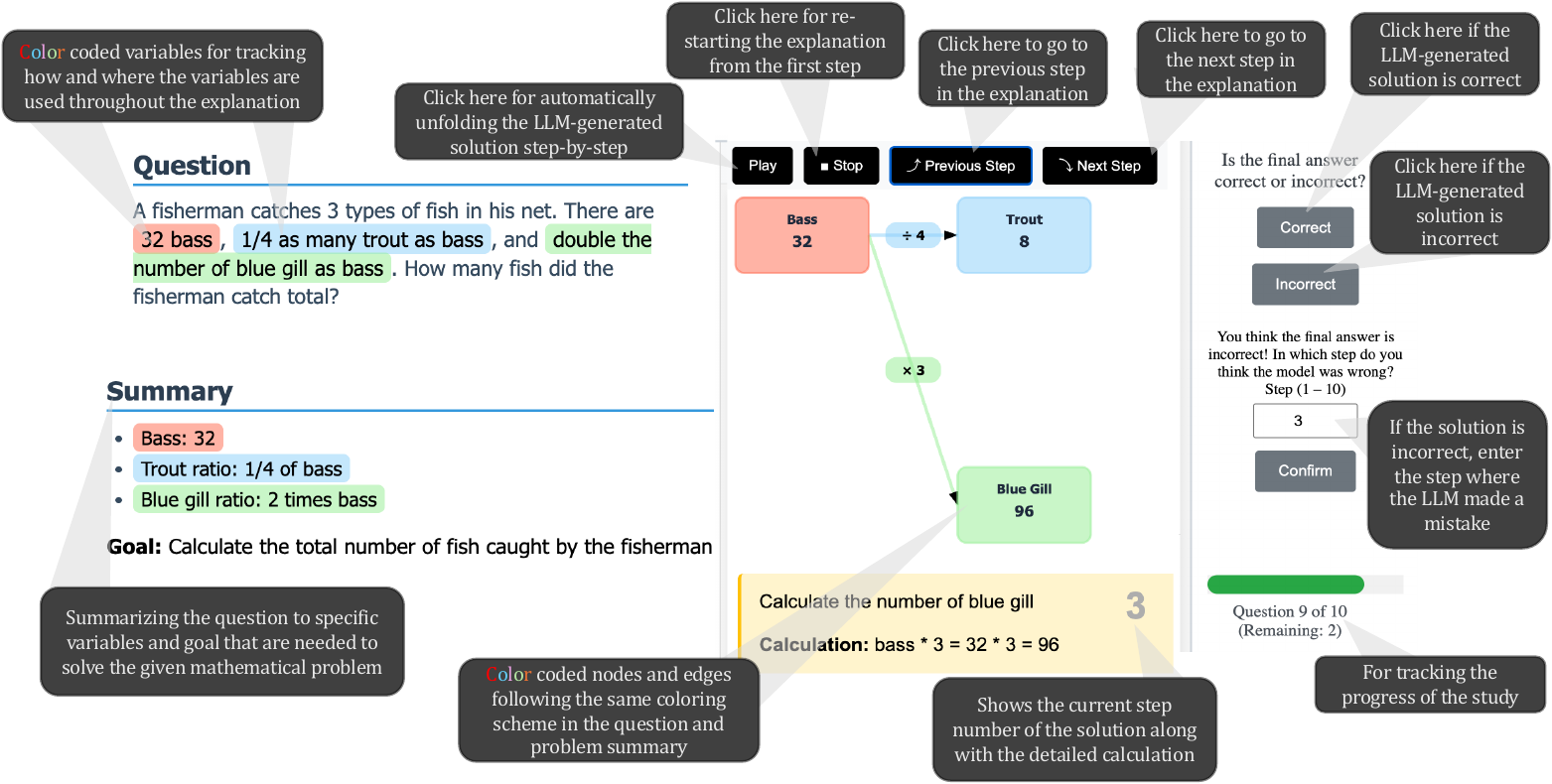}
    \caption{\looseness=-1 Our proposed interactive interface (using iGraph as an example) shows reasoning as connected nodes and edges, using consistent color-coded variables across the question, summary, and explanation. Interactive buttons let users follow each step, while a verification panel allows them to mark whether the explanation is correct and identify any errors. See Appendix \ref{appendix:display} for more examples of Chain-of-Thought (CoT), Interactive Program-of-Thought (iPoT), Interactive Chain-of-Thought (iCoT) and IGraph explanations. The live demo can be found here: \href{https://huggingface.co/spaces/Miles1999/interactive-llm-xai-TEST}{https://huggingface.co/spaces/Miles1999/interactive-llm-xai-TEST}}
    \label{fig:interface}
\end{figure*}
We conduct a large-scale user study to evaluate how different explanation formats influence interpretability, error detection, and engagement when reviewing LLM-generated math explanations. The study follows a between-subjects design, where each participant was randomly assigned to one of four interface formats (CoT, iCoT, iPoT, or iGraph). IGraph interface is displayed as an example in Fig. \ref{fig:interface} with detailed explanation for each interface design features.  Each participant reviews ten explanations presented in their assigned interface format. The set of explanations contains both correct and incorrect solutions to grade-school math problems from the GSM8K dataset, with errors systematically injected. One hundred and twenty-nine undergraduate students ($n = 129$) from two U.S. universities are recruited through class announcements. We acknowledge that GSM8K problems are designed for elementary and middle school students but our user study only includes undergraduate participants. The justification for our design choice are discussed in the Appendix. \ref{appendix:mismatch}.  After the experiment, we plot a histogram showing how long participants take to answer each question. Data from four participants are excluded because their completion times are either unusually short or excessively long (average completion time for each question is less than 20 seconds or greater than 400 seconds), suggesting that they may have randomly guessed the answer or become distracted. Including these cases can bias both the time and the accuracy results. After the screen, the results from 125 participants are used for later analysis. 
% The participant pool was diverse in terms of major and academic background. 
All participants have completed at least one mathematics course beyond high school. Participation is voluntary, and students are informed that their answers will remain completely anonymous. 
% The study was approved by the Institutional Review Board (IRB #2018-15).~\chirag{Add a line about the IRB}

\subsection{Setup and Procedure}

The study is conducted in a controlled classroom setting. Each participant accessed the experiment through an external link and was provided with both written and verbal instructions. The experiment is deployed on HuggingFace Spaces, which hosts both the frontend and backend within a unified environment. This cloud-based setup allows participants to run the study directly in a standard web browser without requiring any local installation or configuration.

\looseness=-1 On the front end, we let LLM pre-generate a total of $2000$ explanation interfaces in HTML format, which corresponds to 500 explanations in each of the four interface formats. These HTML interfaces are then stored on HuggingFace and accessed during the study. A separate experiment wrapper is implemented in HTML, which embeds the explanation interface files through an iframe. This wrapper provides the environment in which participants complete their tasks, and it contains additional experimental features. Specifically, the wrapper presents the (in)correct buttons for judging correctness. A text box is also added for participants to input the specific step at which the error occurs. A progress bar is implemented to inform participants of the number of explanations they have completed, helping them keep track of their progress during the experiment. On the backend, HuggingFace records participants’ responses and interaction data. Each submission is anonymized and stored in a HuggingFace Dataset, which also contains automatically logged statistics such as verification accuracy, time spent on each question, and the full click history of button interactions. These logs are then converted into structured JSON files and stored within the dataset for later analysis. 

\looseness=-1\noindent\xhdr{Experiment} The experiment contains three stages. First, participants are introduced to the study and informed of its goals, procedures, and their right to withdraw at any time without penalty. Consent is obtained before proceeding. Each participant is then randomly assigned to one of the four explanation formats and shown a sequence of ten explanations drawn from the LLM pre-generated explanation pool. Each sequence contains one correct explanation and nine erroneous explanations, each corresponding to a distinct error type. For each explanation, participants are asked to determine its correctness, and if incorrect, to identify the specific error step. Participants can use the playback controls to revisit previous steps within a single explanation, but once their judgment is submitted, they can not return to revise their answers. There is no time limit imposed on individual questions. Finally, after completing all ten explanations, participants are asked to complete a brief survey reflecting on their overall experience, including ratings of clarity, usability, and engagement.

\subsection{Evaluation Metrics}
\looseness=-1 We collect quantitative and qualitative measures to capture both performance outcomes and user experiences: \textit{i) Verification Accuracy~\cite{nguyen2024interpretable}:} For each explanation, we record whether participants correctly labeled the solution as correct or incorrect, providing a direct indicator of how effectively each interface supports error detection and overall comprehension; \textit{ii) Error Localization~\cite{kamoi2024evaluating}:} In cases where the user identifies an explanation as incorrect, they are required to indicate the specific step at which the error occurred. This allows us to assess not only error detection but also participants' ability to pinpoint the exact reasoning breakdown (avoiding the user answering by guessing); \textit{iii) Response Time~\cite{nguyen2025hot}:} The system automatically logs the total time spent on each explanation, which we use as a proxy for cognitive effort, with longer times potentially indicating greater difficulty in comprehension or in identifying error; \textit{iv) Interaction Behavior~\cite{ooge2022explaining}:} All interface interactions, including play, pause, step forward, and step back, are recorded at the click level. These logs gave us insights into how participants navigate the explanations; and \textit{v) Subjective Feedback~\cite{kim2023help}:} After completing the tasks, participants rate their assigned interface on dimensions such as clarity, ease of use, and overall satisfaction using Likert-scale questions. They are also invited to provide open-ended feedback about what they found helpful or confusing, helping us understand participants' perceptions of strengths and weaknesses in each format.
% This can give us the insights into whether different formats encouraged distinct patterns of engagement.

\subsection{Post survey Questionnaire}
\looseness=-1 After the experiment, participants will complete 2 questionnaires displayed in Appendix \ref{appendix:questionnaire} to assess both their interaction experience and their perceptions of the explanation quality. The questionnaires are designed to address the study’s research questions regarding interpretability, efficiency, and engagement across different explanation formats. All items use a 7-point Likert scale (1 = Strongly Disagree, 7 = Strongly Agree) to capture nuanced user perceptions. Two types of questionnaires are given: \textit{Design Feature} and a \textit{Post-Study}. The question sets are inspired by established measures of engagement, usability, and interpretability in human-AI interaction research and are adapted for the design features of our interfaces~\cite{rong2023towards, kim2023help, zhou2019mhealth}.
% (for the three interactive formats) (for all four formats)

\xhdr{Post-Study Questionnaire} All participants will complete a post-study questionnaire after finishing all explanation tasks. This evaluates overall satisfaction, perceived clarity of explanations, and perceived learning effectiveness (see Appendix~\ref{appendix:questionnaire}, Figure~\ref{fig:post}). Participants rate more general statements such as: ``\textit{This explanation format helped me understand the reasoning behind the solution}'' or ``\textit{I found the interface engaging to use.}''
For the interactive conditions, the questionnaire also captures perceived smoothness of interaction and cognitive effort compared to traditional text-based formats. The CoT format includes only the post-study questionnaire, as it does not include interactive elements that allow feature-specific evaluation. Responses across all questionnaires are analyzed to understand how different interaction mechanisms affect interpretability, efficiency, and user engagement.

\looseness=-1\xhdr{Design Feature Questionnaires} Participants assigned to one of the interactive formats (iCoT, iPoT, or iGraph) will complete a design feature questionnaire immediately after finishing their tasks. These questionnaires aim to assess user perceptions of key design components introduced in the interactive interfaces, including the dual-panel layout, problem summarization, color-coded variables, and step-by-step playback controls see Appendix~\ref{appendix:questionnaire}, Figure~\ref{fig:design}).
Each feature is evaluated through one targeted question, \textit{e.g.,} participants are asked whether the dual-panel layout helped them maintain context or whether the playback controls support reasoning flow at their own pace. The design feature questionnaire allows us to capture which design aspects contributed most strongly to participants' perceived interpretability and usability.

\section{Result \& Discussion}
Next, we present experimental results for our user studies and address the research questions introduced in Sec.~\ref{sec:intro}.

% \begin{figure*}
%     \centering
%     \includegraphics[width=1\linewidth]{result_plot.pdf}
%     \caption{Overview}
%     \label{fig:result_plot}
% \end{figure*}

\subsection{Quantitative Result}
% Here, we answer the research questions introduced in Sec.~\ref{sec:intro}: \textbf{RQ1)} Do interactive explanation formats improve verification accuracy? \textbf{RQ2)} How do interactive formats influence the time required to verify reasoning?
 % achieves the highest verification accuracy
 
\looseness=-1\xhdr{RQ1) Interactive explanations improve verification accuracy} A Kruskal–Wallis ANOVA is conducted to examine whether verification accuracy differed significantly across the four explanation formats (CoT, iCoT, iPoT, and iGraph). The results indicate a statistically significant difference among the groups (H(3) = 13.12, p = 0.004). 

\begin{figure}
    \centering
    \includegraphics[width=0.6\columnwidth]{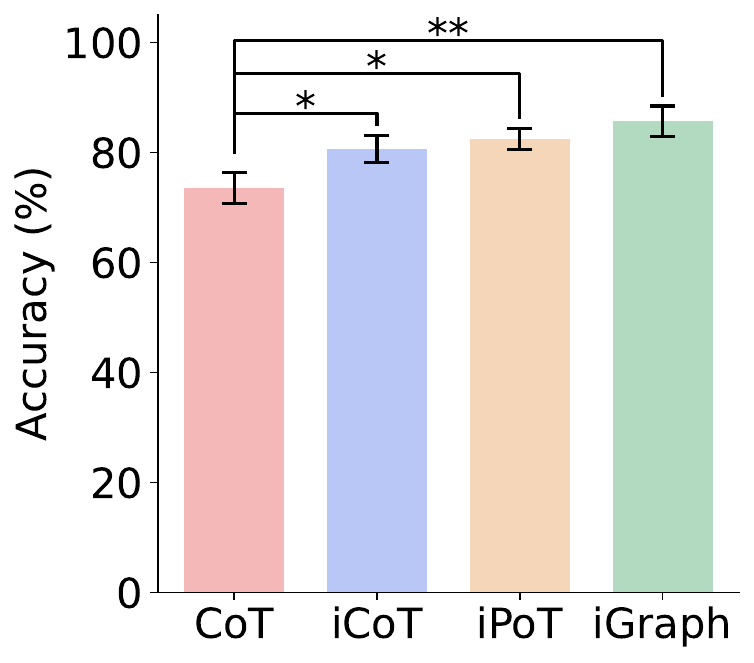}
    % \vspace{-0.25in}
    \caption{\xhdr{Verification accuracy} Results show that participants achieve the highest verification accuracy using iGraph.}
    \vspace{-0.05in}
    \label{fig:verification}
\end{figure}
To further examine differences in verification accuracy between the traditional CoT format and each of the interactive explanation formats, we perform pairwise Mann–Whitney U tests with Bonferroni correction to account for multiple comparisons. The analysis shows that \textbf{iCoT($U = 323.5, p = 0.047, r = 0.32$), iPoT ($U = 306.0, p = 0.023, r = 0.38$), and iGraph ($U = 221.5, p = 0.003, r = 0.51$) all achieve significantly higher verification accuracy compared to the traditional CoT} format. However, differences between the interactive formats themselves (iCoT, iPoT, iGraph) are not statistically significant after Bonferroni correction ($p > 0.05$).
\looseness=-1 Results in Fig.~\ref{fig:verification} demonstrate that interactive explanation formats substantially improve participants' ability to verify reasoning accuracy compared to the traditional CoT. In particular, the iGraph interface achieves the highest verification accuracy ($85.6\%$), followed by the iPoT ($82.5\%$) and iCoT ($80.6\%$) formats, whereas CoT only achieves $73.5\%$. This pattern indicates that interactive explanation formats help users better detect and localize logical faults. These findings align with recent work emphasizing the importance of human-centered explainability, which explains understanding as a process of active answer-seeking rather than passive reception~\cite{kim2024human}. 
% By adding various interactive elements, our interfaces helped users keep track of information more easily, reducing mental effort and making the explanations easier to understand.

\begin{figure}[H]
    \centering

    \includegraphics[width=0.6\columnwidth]{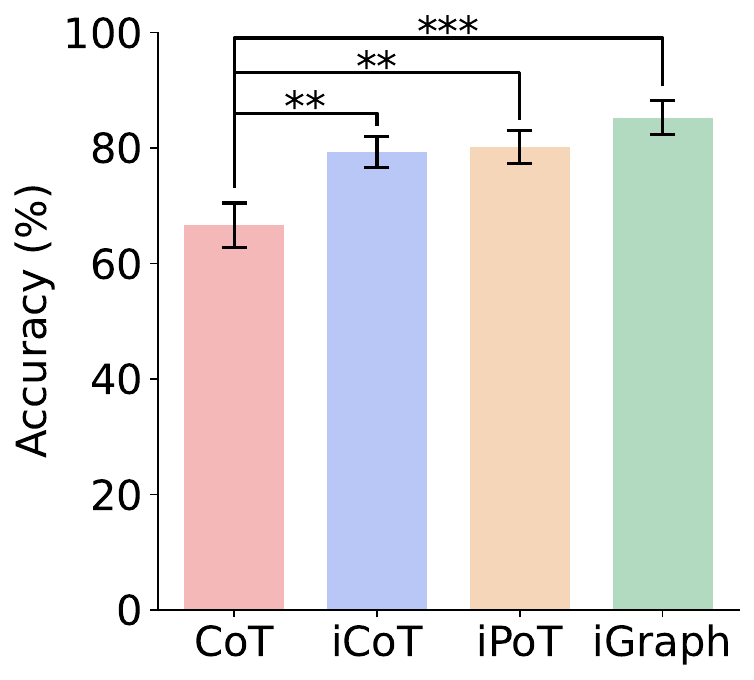}
    \caption{\xhdr{Error localization} Results show that participants were able to detect the exact error step in the LLM's explanation with higher accuracy using iGraph.}
    \label{fig:error}
\end{figure}

\looseness=-1 Next, we use a Kruskal–Wallis ANOVA to examine whether participants' ability to identify the specific step containing an error varies across the four explanation formats. The analysis reveals a statistically significant difference among formats ($H(3) = 13.12, p = 0.004$). Pairwise Mann–Whitney U tests with Bonferroni correction show that \textbf{iCoT($U = 300.8, p = 0.045$), iPoT ($U = 271.5, p = 0.021$), and iGraph ($U = 153.0, p < 0.001$) interfaces all achieve significantly higher wrong-step identification accuracy compared to the traditional CoT} format. Results in Fig.~\ref{fig:error} show that IGraph has achieved the highest wrong-step identification accuracy ($85.2\%$), followed by iPoT ($80.1\%$) and iCoT ($79.3\%$), compared to only $66.1\%$ reached by traditional CoT. The higher wrong-step identification accuracy on iGraph further suggests that interactive explanations helped users understand each step more thoroughly. Prior works in explainable artificial intelligence (XAI) argue that unstructured explanations often overwhelm users with text and often make it hard for users to keep track of each reasoning step~\cite{lakkaraju2019faithful}. Our findings support these arguments by showing that interactive explanations allow users to accurately identify the origin of an error, rather than simply noticing that one exists. Further, visualizing reasoning steps as interconnected nodes and edges transforms purely textual information into a two-dimensional spatial format, making the logic flow easier to follow and understand (this is further verified by our post survey rating results in Fig.~\ref{fig:survey}). 
% Moreover, a study argues that causal transparency (by making the causal relationships more explicit) helps users form more accurate mental models of a system’s reasoning~\cite{shin2021effects}.
% Similarly, visualizing reasoning steps as interconnected nodes and edges transforms purely textual information into a two-dimensional spatial format, which makes the logic flow easier to follow and understand. 
% Prior research has shown that unstructured explanations often overwhelm users with text and often make it hard for user to keep track of each reasoning step\cite{lakkaraju2019faithful}. 
% Our findings support this finding by showing that interactive explanations allows users to better identify the origin of an error, rather than simply noticing that one exists.

In terms of reliance, our results suggest that interactive explanations promote more appropriate reliance on LLM reasoning. The higher verification accuracy (Fig. ~\ref{fig:verification}) and error localization rates (Fig. ~\ref{fig:error}) indicate that participants using interactive formats were better able to critically evaluate the AI's outputs rather than accepting them at face value. Specifically, the ability to identify the exact error step with 85.2\% accuracy (iGraph) compared to 66.1\% (CoT) demonstrates that interactive formats can better support user's critical examination.

\xhdr{RQ2) Participants using interactive explanations respond faster on average, but the time differences are not statistically significant} We also use Kruskal–Wallis ANOVA to see if the time taken by participants to evaluate explanations significantly differed across the four interface formats.

\begin{figure}
    \centering

    \includegraphics[width=0.6\columnwidth]{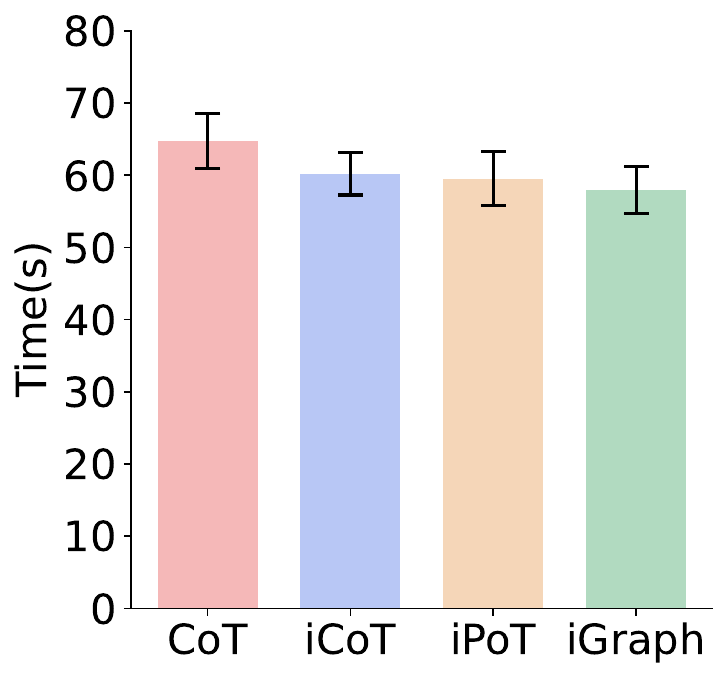}
    \caption{\xhdr{Response time} Results show that participants were able to respond faster using iGraph.}
    \label{fig:response}
\end{figure}
\looseness=-1 The test shows a significant overall difference ($H(3) = 13.12, p = 0.004$), which suggests that the interface type has some effect on how much time participants spend on completing each task. However, the follow-up Mann–Whitney U tests with Bonferroni correction show that none of the pairwise comparisons reach statistical significance. Results in Fig.~\ref{fig:response} show an interesting pattern, where participants assigned to the traditional CoT format spend the most time on each question ($\mu = 64.7$, $\sigma = 40.9$), but participants assigned to the iGraph interface have the fastest response time ($\mu = 57.9$, $\sigma = 41.4$). Participants who are assigned to the iPoT ($\mu = 60.1$, $\sigma = 37.9$) and iCoT ($\mu = 59.5$, $\sigma = 35.8$) interfaces spend slightly less time compared to traditional CoT. This trend hints that interactive formats may have helped participants interpret explanations more efficiently, even if the differences aren't statistically significant. 

Across all quantitative measures, the data show that interactive explanation formats consistently improved verification-related performance and completion time. 
% This result aligns with previous research, which shows that improvements in understanding usually come from making information easier to find and connect[ 4 ]. 
In our interfaces, features such as playback controls and problem summaries allow participants to focus on relevant steps rather than reading the entire explanation.

\subsection{Qualitative Result}
Participants complete a post-experiment questionnaire to evaluate their understanding, engagement, satisfaction, and perceptions of design effectiveness across the four explanation formats. While all four formats are assessed on general measures of comprehension and engagement (G1–G5), only the three interactive formats included additional design-specific questions (D1–D4). Using the responses from these questions, we answer the research question: ``\textit{Do users prefer interactive explanation formats over traditional text-based outputs?}'' 

\begin{figure*}
    \centering
    \includegraphics[width=0.9\textwidth]{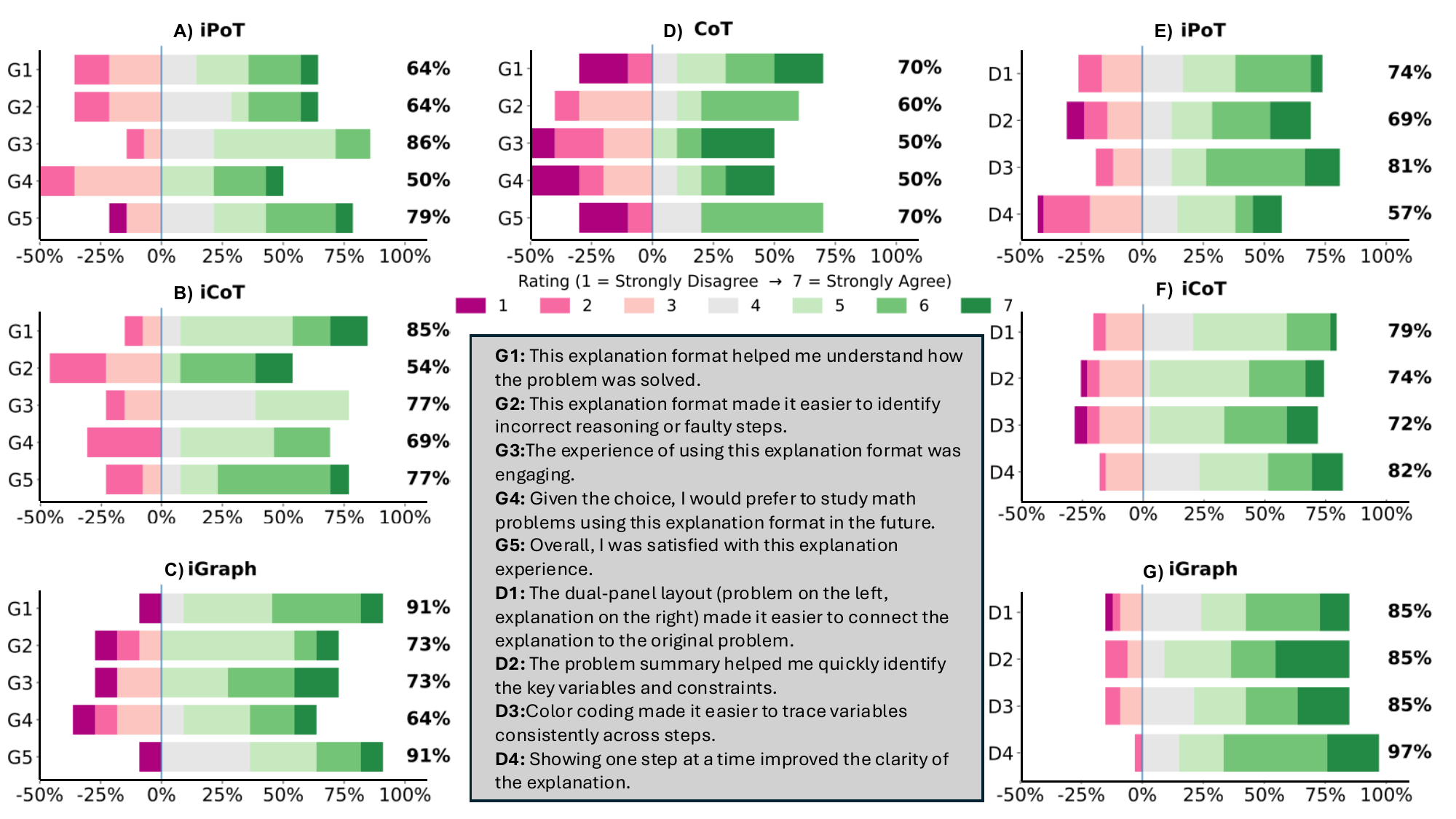}
    \caption{\looseness=-1 Post Survey Questionnaire results for comparing different explanation formats using utility questions (G1-G5) and design-based questions for iPoT, iCoT, and iGraph (D1-D4).  Results from the utility questions show that participants find the interactive formats more effective and engaging, reporting improvements in understanding, error detection, engagement, preference, and overall satisfaction (G1–G5) compared to the traditional CoT. Similarly, results from the design-based questions indicate that participants across all interactive formats consider all four design elements helpful during the experiment. The visualization follows methodology from~\cite{andrews2024aicommentator, andrews2025aimoderator,zhu2023iball}. For the general measures assessment, we find that iGraph (A) achieves the highest rating as compared to CoT (B), iCoT (C).
    }
    \vspace{-0.1in}
    \label{fig:survey}
\end{figure*}

\looseness=-1\noindent\underline{\xhdr{1) General Measures Assessment}} Across the first two questions, assessing the interface's ability to improve \textbf{user understanding (G1)} and \textbf{user ability to identify errors (G2)}, participants consistently rate iGraph the highest (Fig.~\ref{fig:survey}; see plot C), suggesting that the iGraph interface is most effective in helping users understand the explanations and follow the reasoning process. In contrast, both CoT and iCoT receive lower scores on these items, meaning text-based formats are less helpful for understanding and error detection compared to the visual and structured formats(Fig.~\ref{fig:survey}; see plots B and D). 
% A recent study has confirmed that users perceived graph-based explanations as more usable, which led to higher objective understanding \cite{delarue2024evaluating}. 

The third question directly assesses \textbf{self-perceived engagement(G3)} during the experiment. All interactive interfaces achieved strong ratings for engagement (G3) compared to CoT, where a large proportion of participants gave low ratings. The results show that iPoT receives the highest engagement rating, followed by iCoT and iGraph, suggesting that participants find the structured, code-like format of iPoT more engaging(Fig.~\ref{fig:survey}; see plot A-C). Its explicit variable tracking and step-by-step logic may have helped users feel more in control of the reasoning process, which is consistent with research discussing that structured and rule-based explanations promote active engagement and comprehension~\cite{liao2021human}. 
 % These findings also align with the study, which claims that effective explanations should allow users to actively explore reasoning, not just passively consume it~\cite{kim2023help}. 
 Our interfaces foster engagement by allowing users to control reading pace with interactive elements such as playback control. 
 
\xhdr{RQ3) Users prefer interactive interfaces (iGraph, iPoT, iCoT) over the traditional CoT format}  \looseness=-1 The last two questions measure \textbf{future use preference (G4)} and \textbf{overall satisfaction (G5)}. Overall, participants give iGraph the highest satisfaction, while iCoT is the preferred explanation format to study math(Fig.~\ref{fig:survey}; see plot B and C). This pattern indicates that the \textbf{interactive and visually organized formats are perceived as more engaging and enjoyable learning experiences} compared to the static CoT baseline(Fig.~\ref{fig:survey}; see plot D). While iPoT receives high engagement scores, its syntactic structure may have introduced additional interpretive effort for some participants who are not familiar with interpreting coding-based explanations(Fig.~\ref{fig:survey}; see plot A). In contrast, participants consistently rate the traditional CoT interface \textbf{lowest} across both questions, suggesting that long textual explanations are less effective at interpreting math reasoning problems.

\noindent\looseness=-1\underline{\xhdr{2) Design Feature Evaluation}} Participants who interact with iGraph, iPoT, and iCoT also rate four design-specific features: \textbf{dual-panel layout (D1)}, \textbf{problem summarization (D2)}, \textbf{color-coded variables (D3)}, and \textbf{step-by-step presentation (D4)}. The design feature evaluation (D1–D4) is developed to isolate and evaluate how specific interactive design elements within the explanation interfaces contributed to users' comprehension and engagement.

% (Fig.~\ref{fig:survey}; see plot B)
For D1 (dual-panel layout), most participants think that viewing the problem on the left while seeing the explanation on the right makes it easier to connect reasoning steps to the original problem. Participants who are assigned to the iGraph interface show the strongest positive sentiment for this design, with 85\% of participants giving neutral or positive ratings, followed by iCoT with 79\% and iPoT with 74\% (Fig.~\ref{fig:survey}; see plot E-G). This dual representation reduces the extraneous load by \textbf{eliminating the need to scroll and reorient}. Our design aligns with the results from Sweller's study, which claims that working memory has a limited capacity and instructional designs should aim to minimize unnecessary mental effort~\cite{sweller1988cognitive}.  

In D2 (problem summarization), most participants find that concise summaries helped them identify critical variables and constraints before reading the explanation. The interface allows users to focus on understanding the reasoning rather than recalling what each variable represents. iGraph interface has received the most positive rating with $85\%$ of participants giving a neutral or positive rating, followed by iCoT with $74\%$, and iPoT with $69\%$ (Fig.~\ref{fig:survey}; see plot E-G). This design is inspired by a recent study, which suggests that providing learners with key information in advance helps them form a mental model before engaging with complex reasoning~\cite{fenteng2023online}.
% \chirag{Multiple citation of previous works in the results section degrades the novelty of our findings. I would suggest adding qualitative feedback here.} 
% By presenting a problem summary before the step-by-step explanation, the interface reduces cognitive load and allows users to focus on understanding the reasoning rather than recalling what each variable represents.

\looseness=-1 In D3 (color coding), participants reported that consistent color coding across panels helped them trace variables throughout multi-step reasoning processes. iGraph interface received the most positive feedback on this feature, with $85\%$ participants giving neutral or positive rating, followed by iPoT with $81\%$ and iCoT with $74\%$ (Fig.~\ref{fig:survey}; see plot E-G). This feature draws inspiration from a recent study, which links reasoning steps to supporting facts through consistent highlighting~\cite{nguyen2025hot}. Each variable in our interface is assigned a distinct color that persists across the problem statement, summary, and explanation. This visual consistency aims to help users trace relationships between inputs and reasoning steps.

Finally, D4 (step-by-step presentation) also received uniformly high ratings across three interactive interfaces. iGraph again has received the most positive rating with $97\%$ participants giving neutral or positive rating, followed by iCoT with $82\%$ (Fig.~\ref{fig:survey}; see plot F and G). However, only $57\%$ of participants who are assigned to iPoT give a neutral or positive rating (Fig.~\ref{fig:survey}; see plot E). This step-by-step reveal design is inspired by the principle of progressive disclosure, which argues that gradually exposing information helps reduce cognitive load and avoid overwhelming users~\cite{springer2018progressive}. In the XAI literature, interactive and incremental explanation designs are recognized as more human-friendly than monolithic explanation dumps, as they allow users to control over depth and pacing~\cite{spillers2010progressive}.

\section {Conclusion and Future Work}
Traditional CoT explanations in LLMs often appear as long blocks of text that make it difficult for learners to trace reasoning or identify mistakes. To address this limitation, we develop three interactive explanations--iGraph, iPoT, and iCoT--to improve the interpretability and engagement of LLM explanations for math problem-solving. 
% and compared them against the traditional CoT baseline. 
% Our study introduced and evaluated three interactive explanation formats designed to improve the interpretability and engagement of LLM explanations for math problem-solving. 
Through a large-scale user study with 125 participants, we find that interactive formats significantly improve users' ability to verify explanations and locate reasoning errors. Overall, the graph interface achieves the highest verification accuracy, which indicates that visually structured reasoning supports a more intuitive understanding of logical dependencies. 
% The iPoT and iCoT interface also improved readability with interactivity. 
% These findings demonstrate that structured and interactive explanation formats can meaningfully aid comprehension and reduce cognitive load compared to static textual outputs.
Our work provides an empirical foundation for designing interactive reasoning systems that foster deeper engagement and better understanding of AI-generated explanations. However, the study also highlights trade-offs such as visual complexity in graphs, syntactic demands in code, and sequential navigation in text, all of which require careful design calibration. Future research should explore adaptive explanation systems that personalize interactivity based on users' cognitive preferences and learning styles.

\section{Limitation}
We acknowledge the imbalanced 1:9 ratio of correct to incorrect responses in our dataset. This design choice prioritize experimental feasibility and data quality over perfect balance. A balanced design would have required 18 questions (9 correct, 9 incorrect), which we think is infeasible given our goal of maintaining participant focus and avoiding cognitive fatigue. Prior research indicates that prolonged verification tasks lead to declining attention and accuracy \cite{zumbach2006cognitive, krosnick1991response, kost2018impact}. We therefore limited the study to 10 questions, designed to take 20-30 minutes, which allowed us to cover all 9 error categories while ensuring participants maintained engagement throughout the experiment.

\section{GenAI Usage Disclosure}
Generative AI tools were used during this research in a limited and well-documented manner. Specifically, Anthropic’s Claude 2.8 was used to assist in generating the interactive explanation interfaces from structured prompts and templates. Additionally, OpenAI’s GPT-5 was used selectively to polish phrasing and improve readability in sections of the manuscript. No generative AI models were used to produce or alter empirical data, statistical analyses, or study results.

\begin{acks} 
We thank all the participants in our user study. GN was supported by an Auburn University Presidential Graduate Research Fellowship.
AN was supported by NSF Grant \#2145767, NAIRR award \#240116, donations from NaphCare Foundation, and research gifts from Adobe Research and Google. C.A. is supported, in part, by grants from Capital One, LaCross Institute for Ethical AI in Business, the UVA Environmental Institute, OpenAI Researcher Program, Thinking Machine's Tinker Research Grant, and Cohere. The views expressed are those of the authors and do not reflect the official policy or the position of the funding agencies.
\end{acks} 

% \newpage
\bibliographystyle{ACM-Reference-Format}
\bibliography{references}

\appendix

\section{Interface Displays}
\label{appendix:display}
\begin{figure*}
    \centering
    \includegraphics[width=0.9\textwidth]{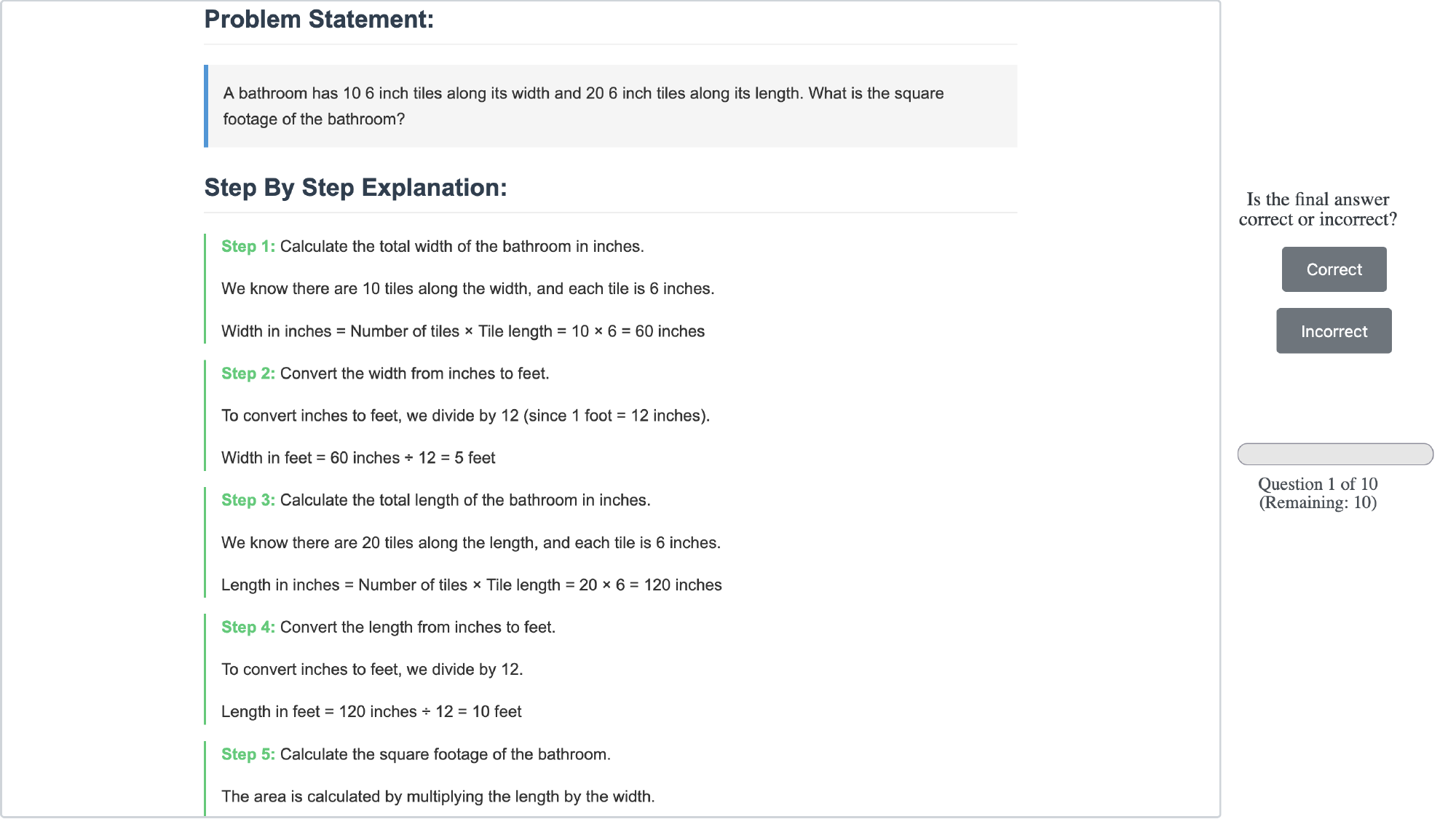}
    \caption{\looseness=-1\textbf{CoT interface }}
    \label{fig:cot}
\end{figure*}

\begin{figure*}
    \centering
    \includegraphics[width=0.9\textwidth]{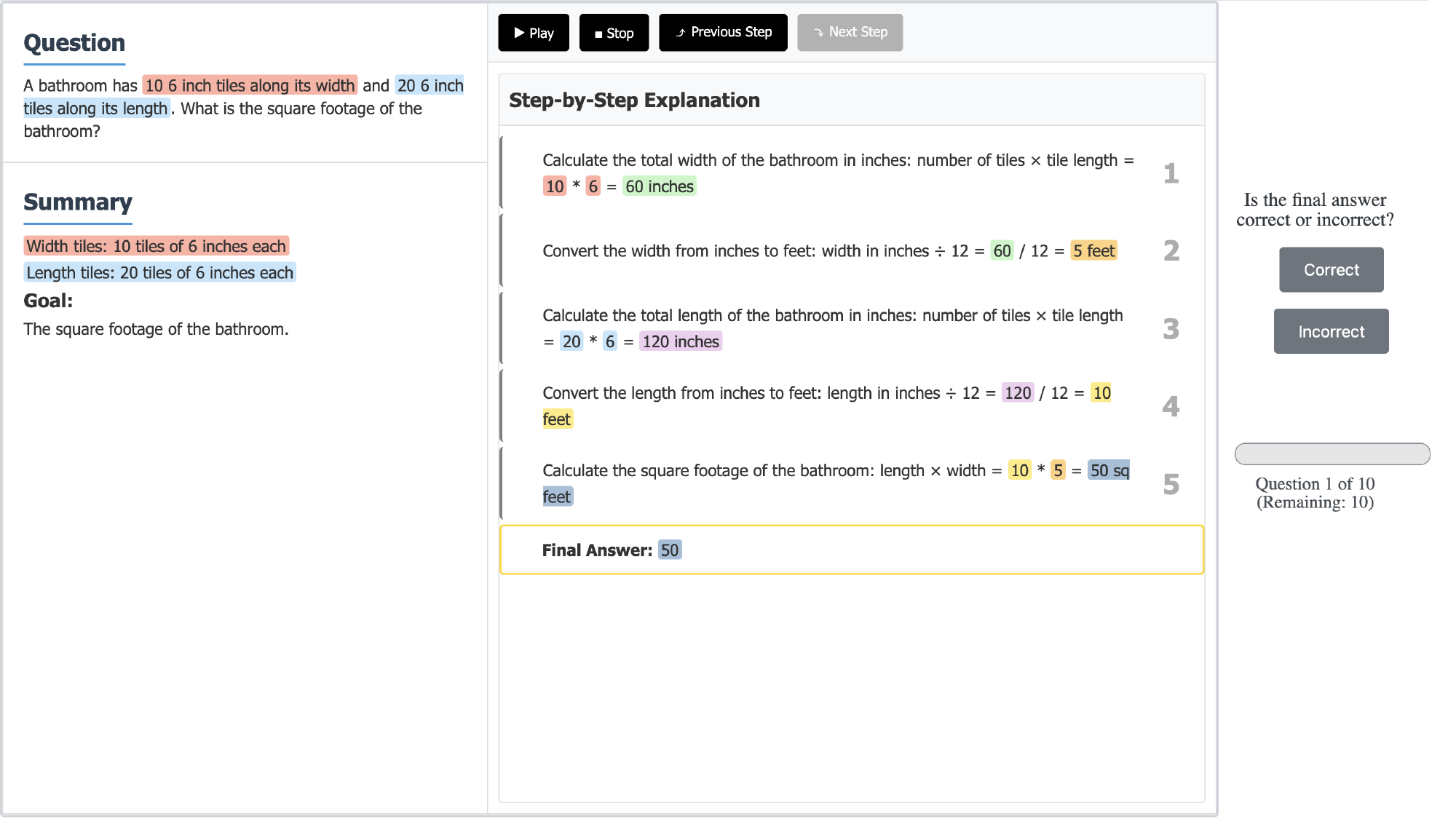}
    \caption{\looseness=-1\textbf{iCoT interface }}
    \label{fig:icot}
\end{figure*}

\begin{figure*}
    \centering
    \includegraphics[width=0.9\textwidth]{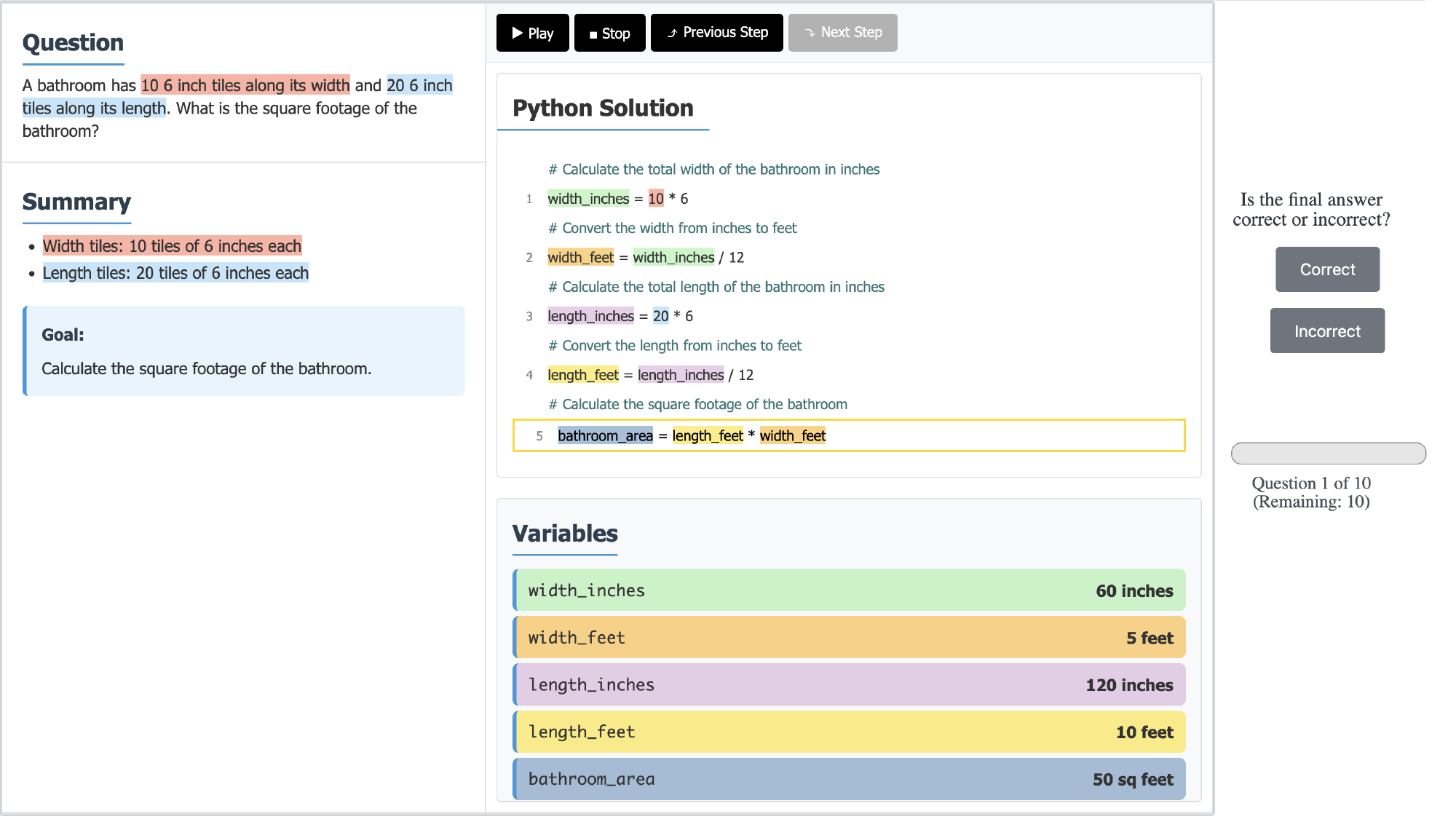}
    \caption{\looseness=-1\textbf{iPoT interface }}
    \label{fig:ipot}
\end{figure*}

\begin{figure*}
    \centering
    \includegraphics[width=0.9\textwidth]{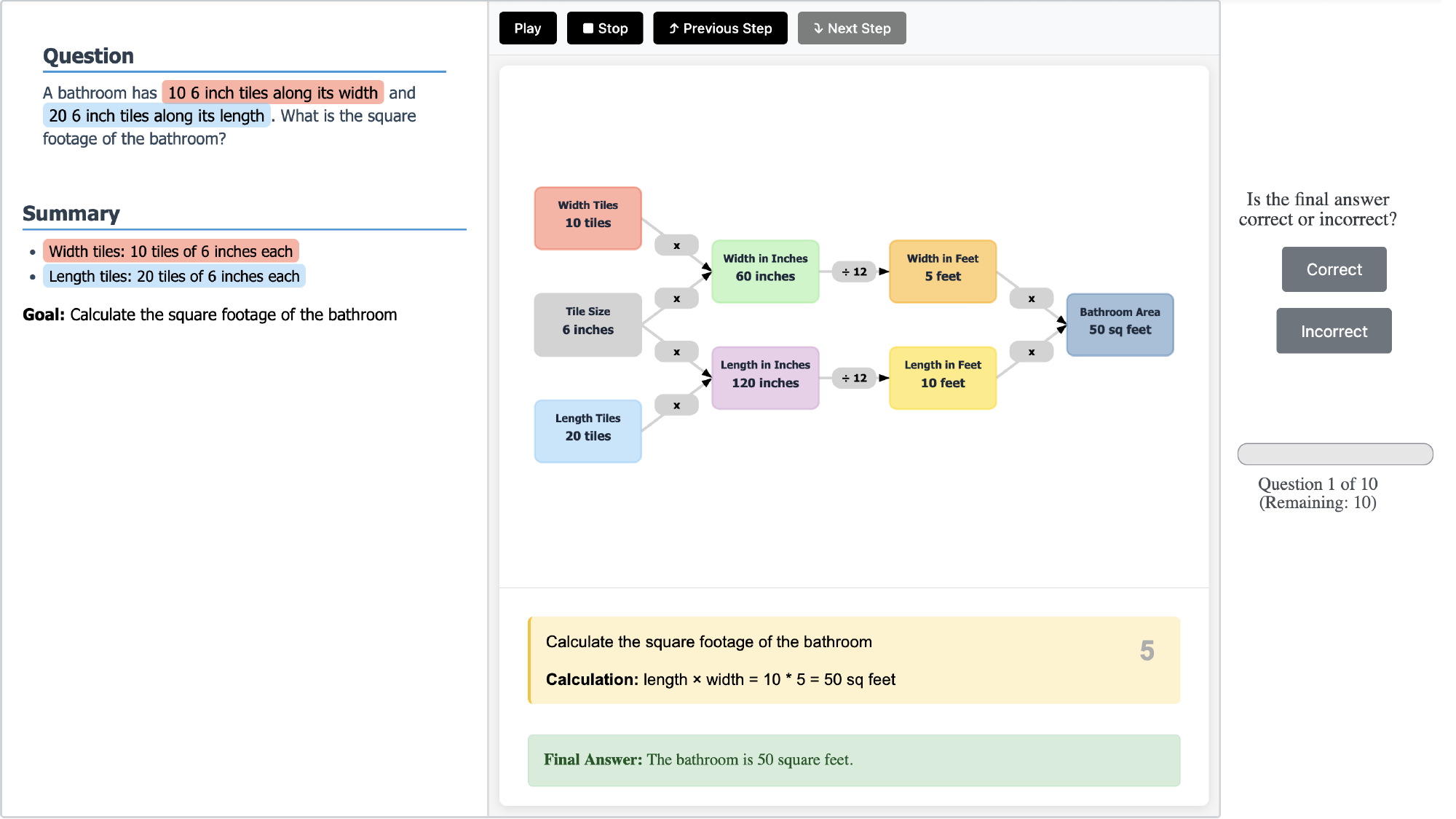}
    \caption{\looseness=-1\textbf{iGraph interface }}
    \label{fig:igraph}
\end{figure*}

This appendix presents examples of all four explanation interface formats evaluated in our study: iGraph (Fig.\ref{fig:igraph}), iCoT (Fig.\ref{fig:icot}), iPoT (Fig.\ref{fig:ipot}), and CoT (Fig.\ref{fig:cot}).
\section{Questionnaire Display}
\label{appendix:questionnaire}
\begin{figure*}
    \centering
    \includegraphics[width=0.7\textwidth]{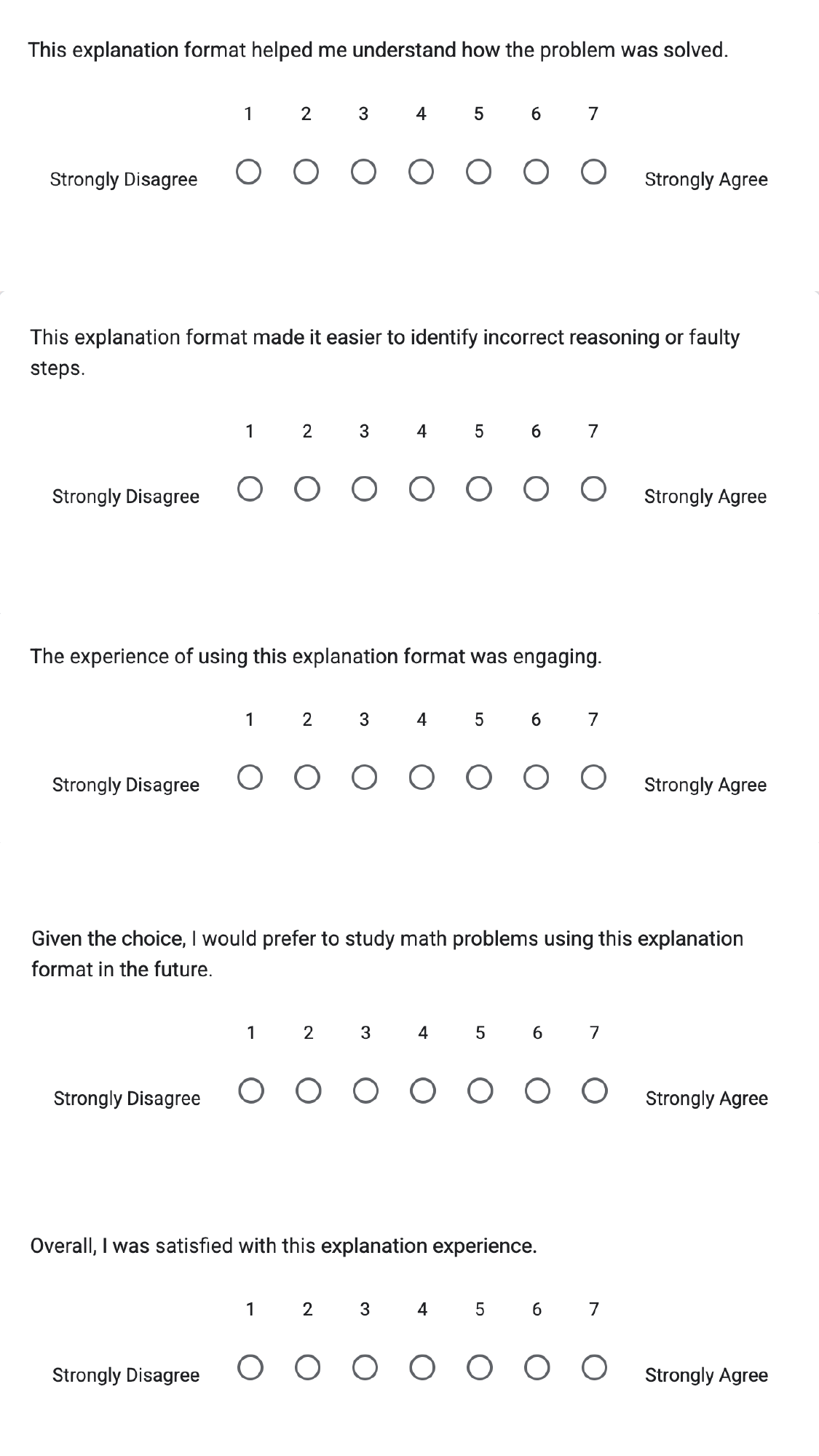}
    \caption{\looseness=-1\textbf{Post-Study Questionnaire}}
    \label{fig:post}
\end{figure*}

\begin{figure*}
    \centering
    \includegraphics[width=0.7\textwidth]{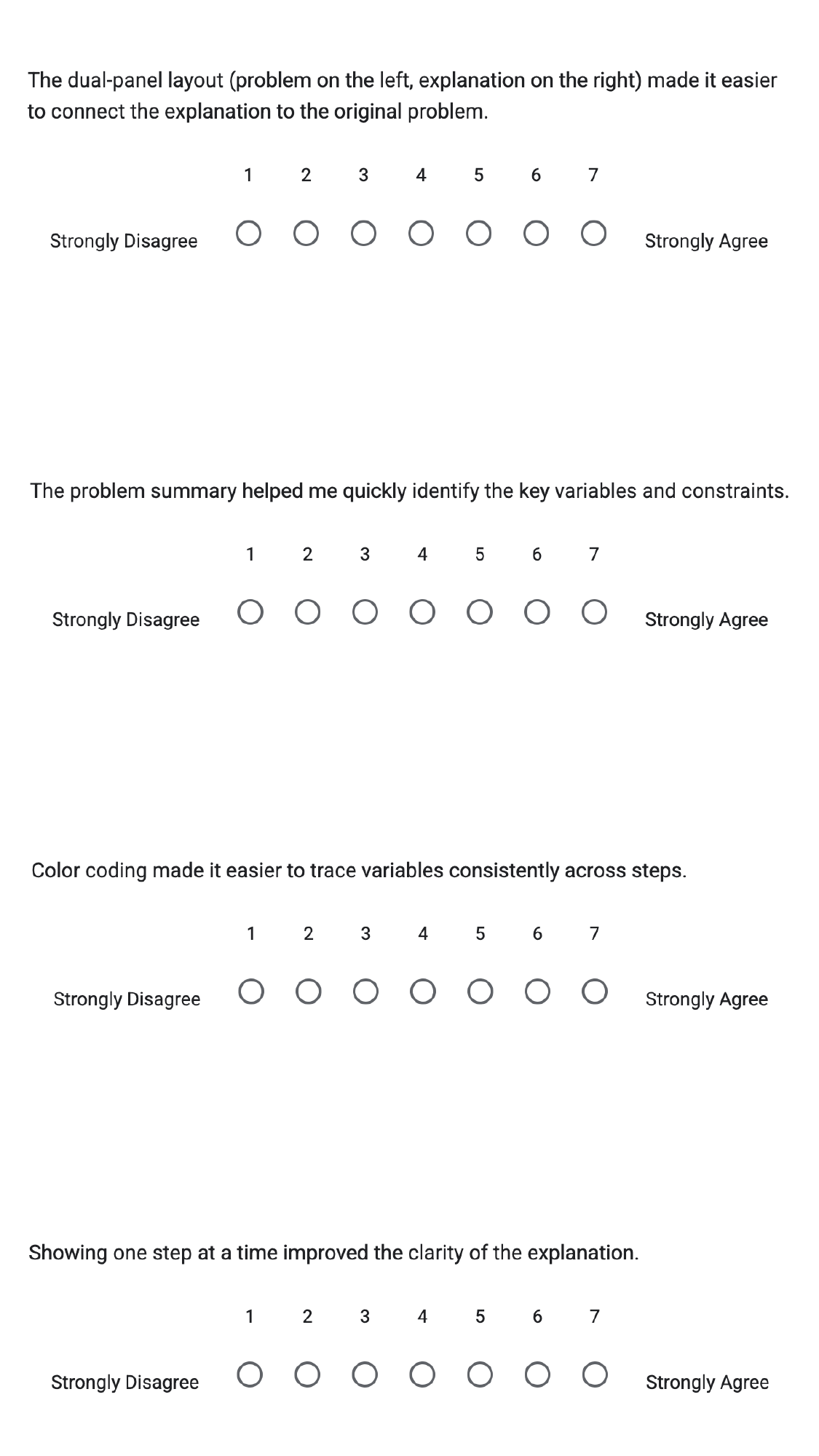}
    \caption{\looseness=-1\textbf{Design Features Questionnaire}}
    \label{fig:design}
\end{figure*}

This appendix presents two post-study questionnaires used in the experiment. The Post-Study Questionnaire (Fig.\ref{fig:post}) focuses on assessing overall engagement and satisfaction with the explanation format, while the Design Features Questionnaire (Fig.\ref{fig:design}) focuses on evaluating specific design features of the interactive interfaces.


\section{Verification Accuracy By Error Type}
\label{appendix:error_category}
\begin{figure*}
    \centering
    \includegraphics[width=1\textwidth]{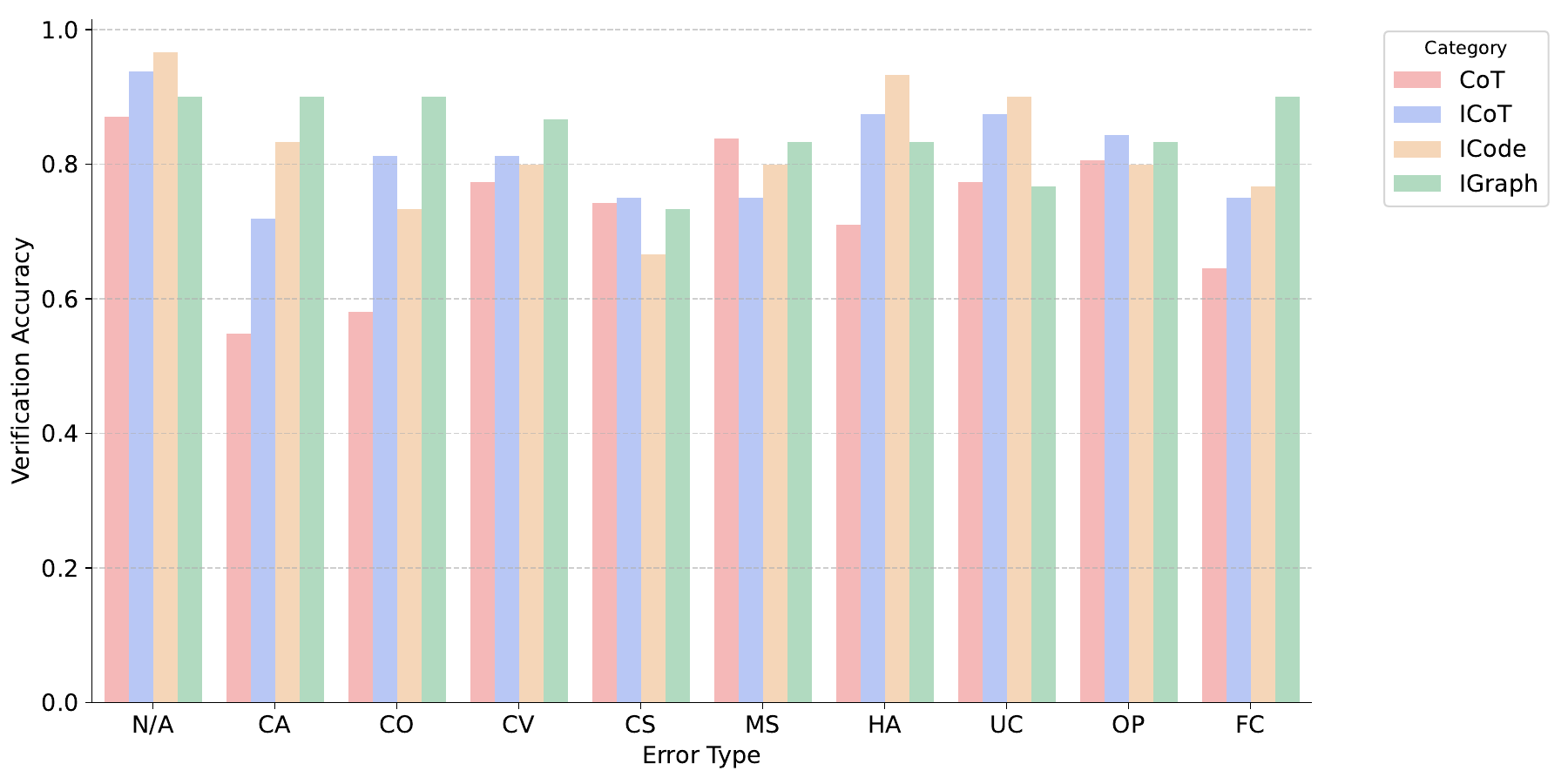}
    \caption{\looseness=-1\textbf{Verification Accuracy By Error Categories} Verification accuracy across the nine error categories and correct explanations (N/A). Interactive formats (iCoT, iPoT, iGraph) consistently show numerically higher accuracy compared to traditional CoT across most error types, though differences within individual categories are not statistically significant.}
    \label{fig:verification_error}
\end{figure*}

To examine whether certain categories of errors are more easily detected across different interface formats, we analyzed verification accuracy for each of the nine error types used in our study. For each error category, we performed a Kruskal–Wallis ANOVA to test whether verification accuracy differed significantly across the four explanation formats (CoT, iCoT, iPoT, and iGraph).
The analysis reveals that verification accuracy did not differ significantly across interface formats for any of the nine error types (all p > 0.05). Figure. ~\ref{fig:verification_error} shows the verification accuracy across all four interface formats for each error category. While no individual error type shows statistically significant differences, the pattern suggests that interactive formats (iCoT, iPoT, iGraph) tend to achieve numerically higher accuracy compared to traditional CoT across most error categories. These findings indicate that the benefits of interactive explanations generalize across different types of logical errors rather than being specific to particular error categories.

\section{Participant-Task Mismatch and Generalizability}
\label{appendix:mismatch}
We acknowledge that GSM8K problems are designed for elementary and middle school students (grades 3-8), with problems typically requiring 5-10 minutes for students still learning these mathematical concepts to solve. Our participants are undergraduate students and above, which creates a mismatch in mathematical sophistication. However, this design choice reflects our study's focus on verification capability rather than problem-solving ability. Specifically, our study examines how effectively humans can verify LLM-generated reasoning chains, which is a cognitively distinct task from solving problems independently. Participants serve as verifiers (analogous to code reviewers), while LLMs serve as solvers (analogous to code authors). We investigate which interface designs best help human verifiers to detect errors in multi-step reasoning produced by AI systems, which represents the realistic use case where humans oversee AI-generated outputs.

While our results primarily reflect interface effects on error detection for users who understand the underlying content, the interface design principles we identify are broadly applicable across domains. The most successful interfaces from our study can transfer to other verification contexts requiring human oversight of LLM reasoning, such as spreadsheet analysis and medical diagnosis \cite{nguyen2024interpretable, le2025s}. Our findings provide generalizable insights into how interactive explanations support human-AI verification across reasoning-heavy domains.

\end{document}